\documentclass[pdflatex,sn-mathphys]{sn-jnl}

\input{macros.sty}

\jyear{2023}

\begin{document}

\title{Continuity Equation for the Flow of Fisher Information in Wave Scattering}

\author[1]{\fnm{Jakob} \sur{Hüpfl}}
\email{jakob.huepfl@tuwien.ac.at}
\equalcont{These authors contributed equally to this work.}
\author[1]{\fnm{Felix} \sur{Russo}}
\equalcont{These authors contributed equally to this work.}
\author[1]{\fnm{Lukas M.} \sur{Rachbauer}}
\author[2]{\fnm{Dorian} \sur{Bouchet}}
\author[3]{\fnm{Junjie} \sur{Lu}}
\author*[3]{\fnm{Ulrich} \sur{Kuhl}}
\author*[1]{\fnm{Stefan} \sur{Rotter}}
\email{stefan.rotter@tuwien.ac.at; ulrich.kuhl@univ-cotedazur.fr}

\affil[1]{\orgdiv{Institute for Theoretical Physics}, \orgname{Vienna University of Technology (TU Wien)}, \orgaddress{\city{Vienna}, \postcode{1040}, \state{Vienna}, \country{Austria}}}

\affil[2]{\orgdiv{LIPhy}, \orgname{Université Grenoble Alpes, CNRS}, \orgaddress{\city{Grenoble}, \postcode{38402}, \state{Auvergne-Rhône-Alpes}, \country{France}}}
 
\affil[3]{\orgdiv{Institut de Physique de Nice (INPHYNI)}, \orgname{ Universit\'e Côte d’Azur,CNRS}, \orgaddress{\city{Nice}, \postcode{06108}, \state{Alpes-Maritimes}, \country{France}}}


\abstract{
Using waves to explore our environment is a widely used paradigm, ranging from seismology to radar technology, and from bio-medical imaging to precision measurements. 
In all of these fields, the central aim is to gather as much information as possible about an object of interest by sending a probing wave at it and processing the information delivered back to the detector. 
Here, we demonstrate that an electromagnetic wave scattered at an object carries locally defined and conserved information about all of the object's constitutive parameters. Specifically, we introduce here the density and flux of Fisher information for very general types of wave fields and identify corresponding sources and sinks of information through which all these new quantities satisfy a fundamental continuity equation. We experimentally verify our theoretical predictions by studying a movable object embedded inside a disordered environment and by measuring the corresponding Fisher information flux at microwave frequencies. Our results provide a new understanding of the generation and propagation of information and open up new possibilities for tracking and designing the flow of information even in complex environments.
}

\maketitle

One of the fundamental concepts in electromagnetism is the Poynting vector, introduced by John Henry Poynting \cite{10.2307/109449} now almost 140 years ago to describe the flow of energy carried by electromagnetic radiation. This concept is intrinsically linked to the Poynting theorem, describing a local formulation of energy conservation with numerous applications in science and engineering \cite{slepianEnergyFlowElectric1942,geyiFosterReactanceTheorem2000,hawthorneFlowEnergySynchronous1954,kelley1991poynting,direnzoCommunicationModelsReconfigurable2022a}. Electromagnetic fields do, however, not only carry energy but they also gather information 
about the environment through which they travel or from which they are reflected. This property is heavily used in radar technology \cite{6504845}, imaging \cite{webb2022introduction,doi:10.1021/acs.nanolett.9b01822}, and high-precision experiments \cite{PhysRevLett.116.061102,lazarev2012optical}, where information is acquired through interaction between radiation and matter. One of the most crucial tasks in all of these applications is to collect as much information as possible about an object of interest to estimate its defining features, such as its position \cite{doi:10.1126/science.aak9913,vanPutten:12}, its mass \cite{young2018quantitative}, or its shape \cite{4516860,6504845} in the most precise way possible. 

The fundamental quantity that provides exactly the link between a measurable observable and the ultimate estimation precision is the so-called Fisher information, which quantifies the amount of information a signal contains about a certain parameter of interest \cite{kay1993fundamentals}. Specifically, the Cramér-Rao bound states that the estimation precision is fundamentally limited by the inverse Fisher information. Based on this fundamental insight, the concept of Fisher information has been applied to numerous problems ranging from microscopy \cite{doi:10.1126/science.aak9913,PhysRevLett.113.133902,PhysRevX.6.031033,https://doi.org/10.1118/1.1677252} to complex scattering problems \cite{Bouchetnatphys,PhysRevLett.124.133903,Boffety:08}. To increase the precision of measurements, much effort has been dedicated both to the optimal extraction of information from a given output signal \cite{PhysRevLett.113.133902,PhysRevA.88.040102} and to the optimal shaping of an input field to boost the amount of Fisher information delivered to a detector \cite{PhysRevLett.123.250502,Bouchetnatphys,PhysRevResearch.5.013144}. While the input and output channels of such scattering problems have thus already been studied in terms of their potential to maximize the accessible information, little is known about the creation of Fisher information at an object of interest and its propagation through the environment that this object may be embedded in. 

Here, we open the lid of this black box and investigate how electromagnetic fields actually acquire Fisher information by interacting with a target and how they deliver it to the far field even in complex scattering systems. A key aspect of our proposed theoretical framework is that we assign to the scattered electromagnetic field a local Fisher information content at any point in space, in analogy to the energy density. Correspondingly, we demonstrate that also the flow of Fisher information is described by a vectorial flux field, in analogy to the Poynting vector for the flow of energy. This Fisher information flux originates directly from a target object (acting as an information source), and is conserved during propagation away from this object. Formally, this is expressed in terms of a continuity equation that preserves information in the same way as the Poynting theorem preserves energy. Using this framework, we provide a theoretical and experimental visualization of the Fisher information flow through a complex scattering region. We emphasize the difference of this spatial information flow to earlier studies on the flow of quantum Fisher information between a quantum system and its environment \cite{PhysRevA.82.042103}.

\begin{figure}[h!]
    \includegraphics[scale=0.47,valign=t]{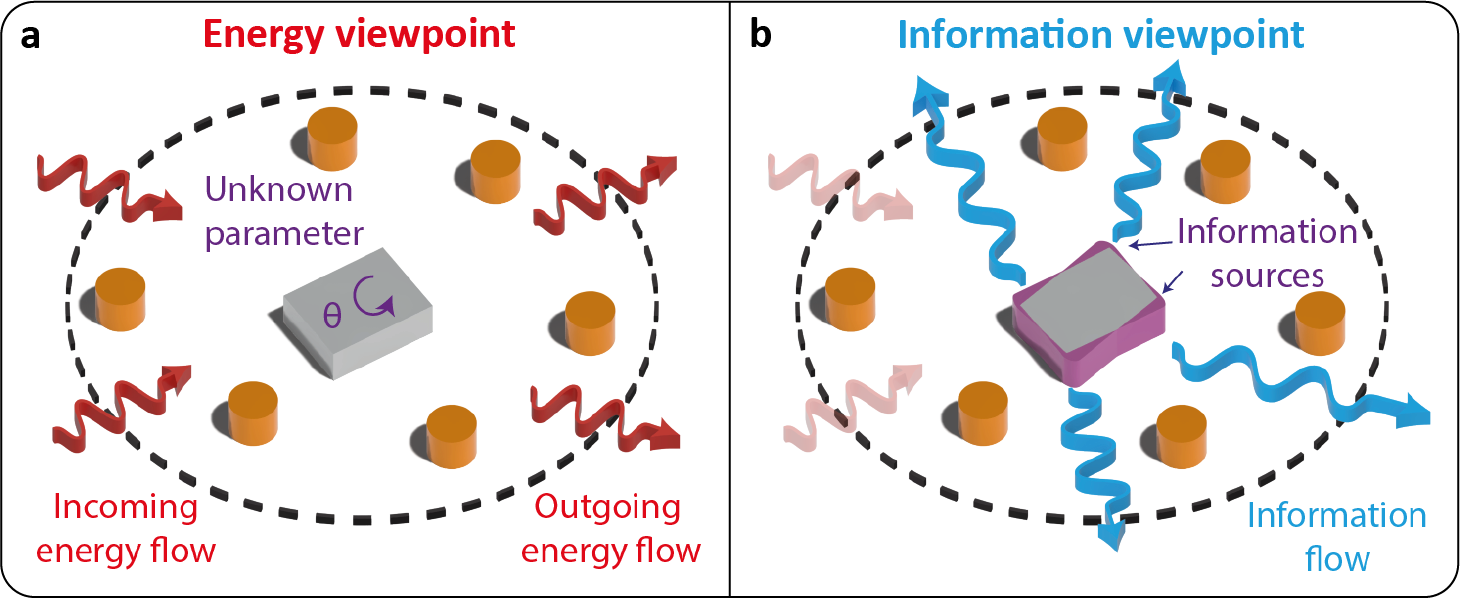}
\caption{\textbf{Illustration of precision measurements in complex systems.} \textbf{a}, In the energy viewpoint, a scattering region containing cylindrical and rectangular scatterers is illuminated by coherent light providing an incoming flow of energy (red incoming arrows). A parameter $\para$ (here the angular orientation of the rectangular scatterer) is estimated from the outgoing energy flow (red outgoing arrows). \textbf{b}, In the alternative information viewpoint presented here, Fisher information on a specific parameter $\para$ is created in those areas that change with varying $\para$ (purple areas at the corners of the rectangle) and the corresponding information flows out of the system (blue outgoing arrows).}
\label{fig:exp_setup}
\end{figure}

\section{Spatial Flow of Information}
To start out, we consider a scattering system made up of dielectric obstacles, which we probe with coherent electromagnetic waves (see Fig.~\ref{fig:exp_setup}\textbf{a} for a sketch of such a setup). The parameter $\para$ characterizes a property of one or several constituents of the considered system, such as the size, shape, or orientation of certain scatterers (see the grey rectangle in Fig.~\ref{fig:exp_setup}\textbf{a}), embedded here inside a disordered system (see orange scatterers). This scenario occurs, for instance, in levitated optomechanics and particle tracking experiments inside a complex scattering environment, where high precision measurements of the particles' properties are necessary for accurate classification and analysis \cite{bon2022some,doi:10.1021/acs.nanolett.9b01822,doi:10.1146/annurev-physchem-050317-021247,doi:10.1126/science.abg3027}. 

Our goal is to estimate a property of the target scatterer parameterized by $\para$ (here the angular orientation of the target in Fig.~\ref{fig:exp_setup}) by illuminating the system with a monochromatic electromagnetic wave (red incoming arrows) and by measuring (part of) the scattered wave (red outgoing arrows).
Labeling the incoming and outgoing energy flux coefficients of the electromagnetic fields as $\vec{c}^{\text{in}}$ and $\vec{c}^{\text{out}}$, respectively, allows us to describe this scattering process through the scattering matrix $\S$, that connects these far field patterns as follows: $\vec{c}^{\text{out}} = \S \vec{c}^{\text{in}}$. The interaction of the field with the system during its propagation imprints information about the local system parameters on the outgoing field. However, the noise inherent in any signal (such as the photon shot noise inherent in coherent electromagnetic waves) limits the amount of information a wave can carry, requiring measurement schemes that maximise the extraction of information \cite{PhysRevA.88.040102,PhysRevLett.123.250502}.

The relevant quantity that measures the amount of information on a specific parameter $\para$ contained in a noisy signal is the so-called Fisher information (FI) $\CFI(\para) = \EX{\{\der[\para] \ln[p(X;\para)]\}^2}$ \cite{kay1993fundamentals}. Here, the noisy measurement data is given by a random variable $X$ with the corresponding probability distribution $p(X;\!\!\para)$ and $\EX$ stands for the expectation value over the noise fluctuations. The significance of the Fisher information is manifested in the Cramér-Rao inequality \cite{kay1993fundamentals}, according to which the inverse of the FI sets a lower bound on the variance of any unbiased estimator $\hat{\para}$ of $\para$ based on the measurement data: $\Var \hat{\para} \geq \CFI(\para)^{-1}$. To obtain high precision estimates, signals with high Fisher information are thus required.

Recent work has shown that in scattering problems the rate of FI, $\CFIr(\theta)$, delivered to the far field by a coherent monochromatic light field can be quantified by the following Hermitian Fisher information operator $\matrix{F} = 4 (\hbar \omega)^{-1} \der[\para] \S^\dagger \der[\para] \S$ \cite{Bouchetnatphys}. Specifically, this operator tells us that an input light field characterized by the coefficient vector $\vec{c}^{\text{in}}$ contains the following rate of FI, $\CFIr(\para)=(\vec{c}^{\text{in}})^\dagger \matrix{F} \vec{c}^{\text{in}}$, that is reachable with a detection scheme based on shot noise limited photodetectors placed in the far field. 
The central question we will address in the following is, how the FI operator can be extended to a local quantity describing the propagation of information in analogy to the energy flow.

To arrive at an expression for a corresponding FI flow from the FI operator $\matrix{F}$, we make use of the analogy to the well-known expressions for energy flow. In the far field, the total outgoing flow of energy is given by the Hermitian matrix $\S^\dagger \S$. This energy flux operator measures the flow of the outgoing wave energy in the asymptotic region as described by the Poynting vector (averaged over a time period)  \cite{jackson1999classical}, 
\begin{equation}
    \poynting = \Re(\Efield_\omega^* \times \Hfield_\omega)/2\,,
\end{equation}
where $\Efield_\omega,\Hfield_\omega$ are the complex amplitudes of the electric and magnetic fields at frequency $\omega$, respectively. To formalize this connection between $\S^\dagger \S$ and $\poynting$, one considers the rate $P$ of energy entering photodetectors of area $A_i$, located in the far field,
\begin{equation}
\label{eq:energyflow}
      P = (\vec{c}^{\text{in}})^\dagger \S^\dagger \S  \vec{c}^{\text{in}}=\sum_i \int_{A_i} \poynting \cdot \d \n_i \,,
\end{equation}
where $\n_i$ are the unit vectors normal to the detectors' surfaces. 

To arrive at a corresponding relation for the rate of FI transfer $\CFIr(\theta)$ (rather than for the energy) into the detectors, we transform the above equation \eqref{eq:energyflow} such as to contain the FI operator (see supplementary material \ref{app:maxFI} for more details),
\begin{equation}\label{eq:fiflow}
     \CFIr(\theta) = 4 (\hbar \omega)^{-1} (\vec{c}^{\text{in}})^\dagger  \der[\para] \S^\dagger \der[\para] \S  \vec{c}^{\text{in}}=(\vec{c}^{\text{in}})^\dagger \matrix{F} \vec{c}^{\text{in}}=\sum_i \int_{A_i} \fiflux \cdot \d \n_i \,.
\end{equation}
The formal analogy between equation (\ref{eq:energyflow}) for the energy and equation (\ref{eq:fiflow}) for the FI, now allows us to identify the FI arriving at the detector surfaces as a corresponding (time averaged) Fisher information flux,
\begin{equation}\label{eq:sfi}
    \fiflux = \frac{2}{\hbar \omega} \Re(\der[\theta]\Efield_\omega^* \times \der[\theta]\Hfield_\omega)\,.
\end{equation}
A sketch of this information flux is shown in Fig.~\ref{fig:exp_setup}\textbf{b} (blue arrows), which is defined by equation (\ref{eq:sfi}) as a vector field assigned to the electromagnetic field not only in the asymptotic region, but in all of space – just like the Poynting vector for energy. In analogy to the Poynting theorem, we can also formulate a continuity equation for the FI flux of the following form,
\begin{equation}
\label{eq:complex_cont_eq}
    \nabla \cdot \fiflux = \fisource\,,
\end{equation}
which now allows us to identify what the sources and sinks of information $\fisource$ are. In the case of static isotropic dielectrics with the permittivity function $\epsilon(r)$, these sources are given by $\fisource = -2 \hbar^{-1} \Im[(\partial_\para \epsilon) \Efield_\omega^* \cdot \partial_\para \Efield_\omega]$. (Similar sources need to be added if free charges or magnetic materials are present, see Supplementary Material~\ref{app:FIField}.) The term $(\partial_\para \epsilon) \Efield_\omega^*$ in this relation for $\fisource$ reveals that the FI is created only in regions where the light field interacts with those portions of matter that vary with perturbations in the parameter $\para$ of interest (purple areas near the corners of the target in Fig.~\ref{fig:exp_setup}\textbf{b}). Note however that this term can not only create but also absorb information that is present in the field similar to how an antenna can absorb or send out electromagnetic waves.

Remarkably, the continuity equation \eqref{eq:complex_cont_eq} encapsulates the fact that the FI flux is a conserved quantity, not only for waves propagating through free space, but also throughout those parts of a medium, which are unaffected by a variation of the parameter $\para$ of interest. In systems with absorption (as modeled by a complex permittivity), not only energy, but also FI gets attenuated by an additional absorption term, $\sigma^{\text{abs}} = -2 \hbar^{-1} \Im(\epsilon) \vert\der[\para] \Efield_\omega\vert^2$. As a consequence, all the FI flux created by information sources either propagates out of the system or gets reabsorbed by information sinks such as system parts that are dissipative or that depend on $\para$.

To locate the sources of FI in a system, it is convenient to introduce a small perturbation to $\theta$ and observe how the scattering landscape changes as a result. For example, in Fig.~\ref{fig:exp_setup}, rotating a target object shows that orientation information is generated most strongly near the corners of that object. However, we highlight that the introduction of such perturbations only serves the purpose of visualising the FI flow and its sources; indeed, even without such perturbations, the scattered wave contains FI for all parameters of a system with which the wave has interacted. To access the Fisher information, an estimator of the parameter $\para$ is required, which comes with the complication that it is generally only valid for a specific system configuration. While an estimator can thus be constructed based on additional system information \cite{doi:10.1126/science.aak9913}, an explicit variation of $\para$ provides direct access to the estimator without any prior knowledge of the system \cite{Bouchetnatphys}.

\section{Experimental Visualisation} 
\label{sec:experiment}
\begin{figure}
\includegraphics[scale=0.17,valign=t]{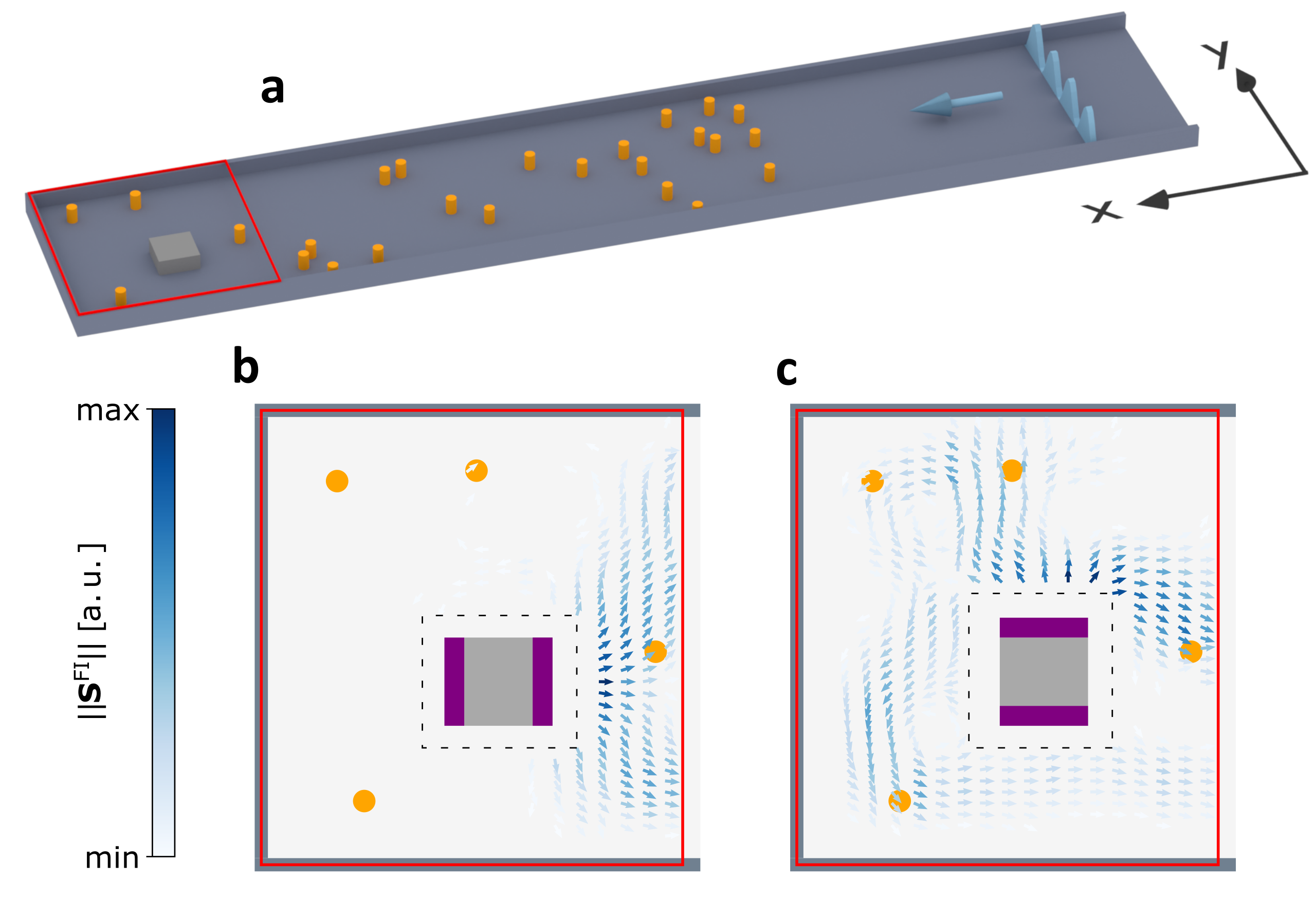}
\caption{\textbf{Fisher information flow in a microwave setup.}  \textbf{a}, Sketch of the experimental setup. We depict the Teflon scatterers as orange cylinders and the metallic target scatterer as a light gray cube. Waves are injected from the right side as indicated by the light blue wavefront. The near field measurement area is highlighted in red. \textbf{b}, Fisher information flux (blue arrows) for the case that the estimated parameter $\theta = \xscat$ is the horizontal position of the grey target object.
\textbf{c}, Fisher Information flux corresponding to its vertical position $\theta = \yscat$. The area around the target is shown for the case that an input state is injected (at $6.45 \GHz$), which maximizes the FI at the output for the given parameter $\theta$. In accordance with our predictions, one clearly observes that the Fisher information emerges from the purple regions where the refractive index changes when $\theta$ is perturbed. The dashed line indicates where the proximity of the target object prohibits near field measurements.}
\label{fig:Exp}
\end{figure}

In order to explicitly demonstrate that the FI flux is also accessible experimentally, we present below corresponding measurements in a microwave setup. The electromagnetic fields are generated and measured here using antennas that give us direct access to the FI flux in the near and far fields. To simplify the setup and reduce complexity, we employ a flat waveguide that is closed by a metal plate on the left side (see sketch in Fig.~\ref{fig:Exp}\textbf{a}). We introduce Teflon scatterers into the system to create a complex scattering geometry. The electromagnetic waves are injected from the open lead on the right (blue wave pattern) at a frequency of $f = 6.45 \GHz$, allowing for 4 propagating transverse modes. We consider the FI fluxes corresponding to the horizontal ($\xscat$) and vertical ($\yscat$) position of a rectangular metallic target scatterer (light gray cuboid), embedded inside a set of 25 small circular Teflon scatterers (orange cylinders). The electric field is measured in the near field around the target (red frame), once before and after shifting the target slightly along the $x$ or $y$ axis, allowing us to construct the FI flux for the relevant parameters $\para=\xscat$ or $\para=\yscat$ using finite differences. A more detailed description of the experimental setup is given in Methods Sec.~\ref{methods:exp_setup}. 

In the experimental results shown in Fig.~\ref{fig:Exp}\textbf{b} we have used the FI operator to inject the optimal incident state that maximizes the collected FI in the far field on the right side of the waveguide \cite{Bouchetnatphys}. The obtained results clearly show how the flux of FI on $\xscat$  (blue arrows) emerges from the target scatterer's right-hand side and propagates toward the far field. This confirms that those parts of the target that change under variation of $\xscat$ (see purple regions in Fig.~\ref{fig:Exp}\textbf{b}) act as sources of FI flux. Moreover, the fact that the state with maximum FI typically originates from the right boundary of the target scatterer (rather than from the left boundary side), can be attributed to the fact that global absorption in the metallic waveguide walls results in a dissipative information loss. Information propagating along the shorter paths emanating from the right target boundary are thus typically more efficient in transferring information to the waveguide output.
For comparison, Fig.~\ref{fig:Exp}\textbf{c} shows the flux of FI corresponding to the vertical position $\yscat$ of the target object. Using again the optimal incident state, we observe that the flux mainly originates here from the upper side of the target object, which constitutes one of the two possible sources of FI (see purple regions in Fig.~\ref{fig:Exp}\textbf{c}).

Despite the structural similarity between the Poynting theorem and our newly introduced FI continuity equation, it would be misleading to assume that the flow of energy and the flow of information exhibit analogous behavior for any given wave field. To provide empirical evidence of this discrepancy, we will simulate a system in which electromagnetic energy is transmitted across the system with high efficiency, while the Fisher information is back-reflected to the source of the probing wave. For this purpose, we open up the left side of the waveguide shown in Fig.~\ref{fig:Exp} and include an additional 25 circular scatterers to the left of our target in our simulations. (For the full geometry see Methods Sec.~\ref{methods:2Dnumerics}.) Previously, we made use of the FI operator $\matrix{F}$ to identify the incident state that transmits the highest FI to the far field. By taking the FI matrices $\matrix{F}_L$ and $\matrix{F}_R$, corresponding to the FI available at the left and right waveguide lead, respectively, we construct the following operator, 
$\matrix{F}^{\text{rel}} = (\matrix{F}_R+\matrix{F}_L)^{-1} \matrix{F}_L\,$. 
This ``relative FI operator'' allows us to identify the state that maximizes the ratio between the information received at the right and the left lead (see Methods Sec.~\ref{app:relFI} for further details).

\begin{figure}
\includegraphics[scale=0.35,valign=t]{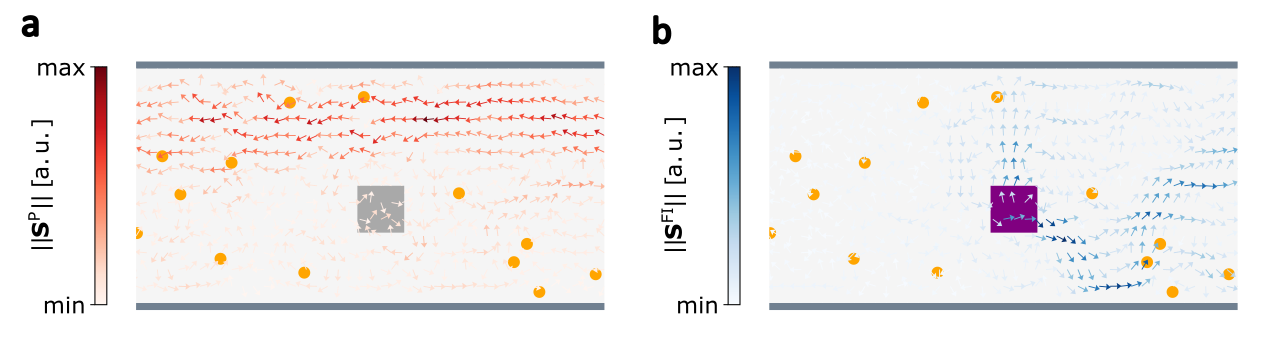}
\caption{\textbf{Comparison of Fisher information and energy flow.} Simulated Poynting vector (red arrows in \textbf{a}), and Fisher information flux (blue arrows in \textbf{b})  around the target scatterer (refractive index $n = 1.44$) at $f = 11.8 \GHz$, when injecting an incident state that maximizes the Fisher information ratio between the transmission and reflection channels. With the estimation parameter $\para$ of interest being here the target scatterer's refractive index, the FI sources are located in its entire area (purple square in \textbf{b}). Both the target scatterer and the disorder (orange circles) have a refractive index of $n=1.44$. We observe that while most of the wave's energy is transmitted ($96.3\%$) to the left (see \textbf{a}), almost all of the Fisher information ($98.4\%$) is flowing to the opposite direction (to the right, see \textbf{b}).
}
\label{fig:relFIflux}
\end{figure}

In Fig.~\ref{fig:relFIflux}\textbf{a} we show the eigenstate of this operator that corresponds to the maximal eigenvalue. The corresponding state features a distribution of Poynting vectors (red arrows), which can be seen to follow a pattern of strong transmission from right to left along the top of the waveguide. 
On the contrary, the FI flux is flowing to the right side of the waveguide and thus in the opposite direction to the dominant flow of energy in this system. 
This not only highlights that the FI flux and the Poynting vector represent distinct concepts, but also that we can collect information about a parameter at a detector (located in the right lead), with very little loss of information to a potential eavesdropper (assumed here to be located in the left lead).

\section{Information density as a local property of electromagnetic waves}

In a next step, we investigate the dynamical process of FI generation at a target object and its storage within the scattered wave. For this purpose, we extend our framework from the time-independent continuity equation (\ref{eq:complex_cont_eq}) to the temporal domain. To be specific, we consider quasi-monochromatic wave packets, denoted as $\Efield = \Re [\Efield_\omega(\vec{r},t) \expo{-\i \omega t}]$, where the complex amplitude $\Efield_\omega$  varies slowly with time and thereby determines the wavepacket's envelope function.
In analogy to the energy density, we introduce the Fisher information density, $\fidensity = (\hbar \omega)^{-1} (\epsilon \vert\der[\para] \Efield_\omega \vert^2 + \mu_0 \vert\der[\para] \Hfield_\omega \vert^2)$, which provides the missing term to complete the fully time-dependent continuity equation, 
\begin{equation}
\label{eq:cont_eq}
\nabla \cdot \fiflux + \der[t] \fidensity = \fisource\,.
\end{equation}
This ``Poynting theorem'' for the FI contains the FI flux, $\fiflux$, and sources, $\fisource$, as defined already for the time-independent equation \eqref{eq:complex_cont_eq}. The newly found FI density, $\fidensity$, describes the local density of Fisher information of the electromagnetic field at any given location.

\begin{figure*}[h!]
\includegraphics[scale=0.85,valign=t]{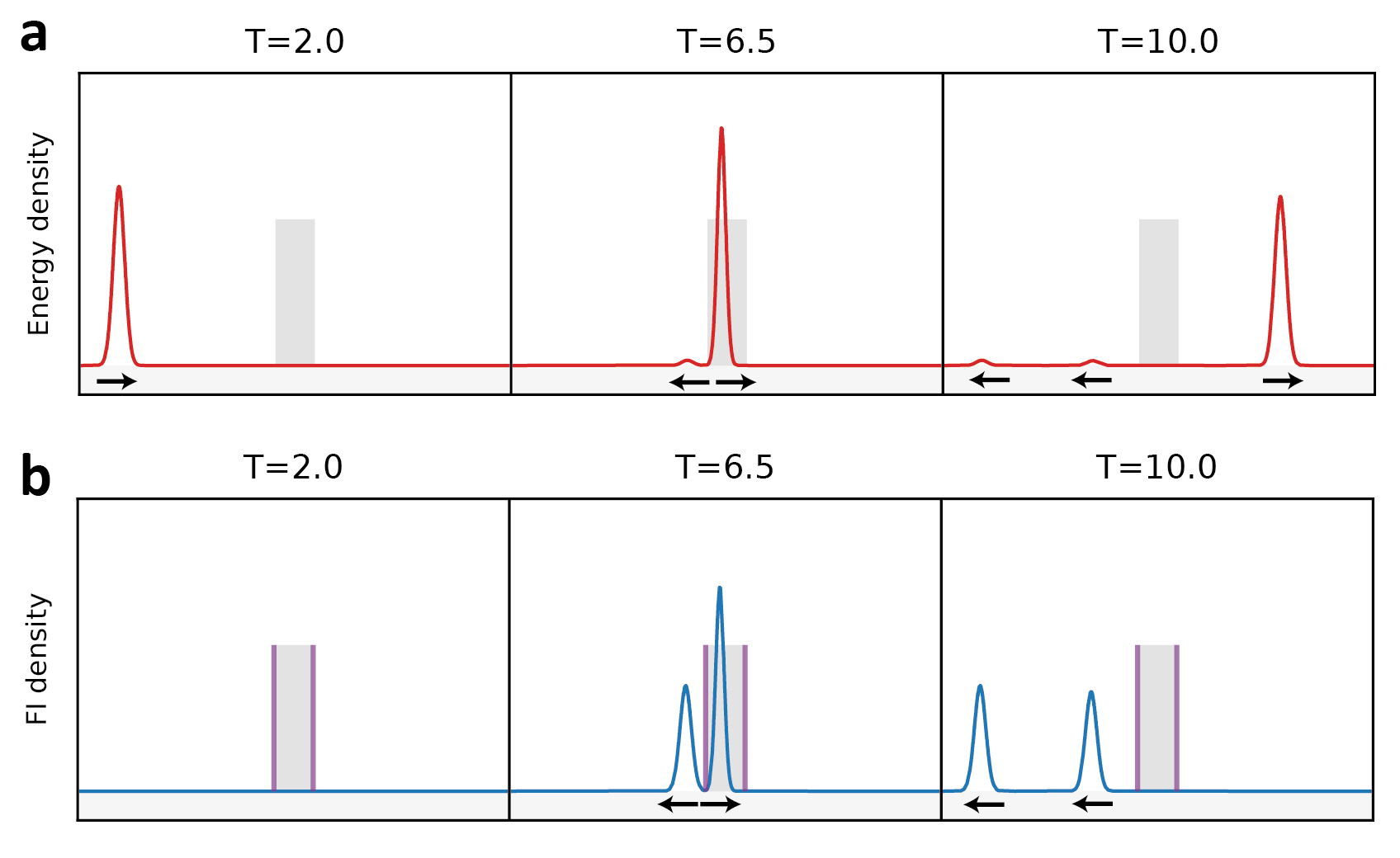}
\caption{\textbf{Fisher information content of a wave packet.} A one-dimensional wave packet closely centered around a frequency $\omega$ scatters off a dielectric (grey rectangle), whose position is the parameter of interest. \textbf{a}, Energy density envelope (red) at times $T=2,6.5,10$ (arbitrary units). The black arrows indicate the direction of motion of the wave packet. \textbf{b}, Fisher information density (blue) at the same times. The information is created at the sources located at the two sides of the scatterer (purple). Most of the energy is transmitted, whereas the information is solely flowing into the reflection channel. 
}
\label{fig:time_dependent}
\end{figure*}

In order to illustrate these quantities, we show in Fig.~\ref{fig:time_dependent} the propagation of a one dimensional wave packet scattering off a barrier (refractive index $n \approx 1.41$). The FI that we consider here, corresponds to the $x$ position of the barrier (grey rectangle), such that the information sources and sinks are located at the very left and right sides of the barrier (purple bars). In panel \textbf{a} of Fig.~\ref{fig:time_dependent} we plot the energy envelope of the wavepacket entering from the left and scattering off the barrier. While a small part of the energy is reflected at the barrier, most of it is being transmitted. 
For comparison, panel \textbf{b} depicts the FI density stored in the wave. First of all, we see that the incoming wave packet does not carry any FI before hitting the barrier. When the wave packet overlaps with the left side of the barrier, two pulses of FI are created (see figure in the middle of panel \textbf{b}). Remarkably, when the same happens at the overlap of the wave packet with the right side of the barrier, the right-moving FI wave packet created at this moment destructively interferes with the right-moving packet created earlier at the left side of the barrier (due to a phase shift of $\pi$). This reflects the fact that the transmitted wave is entirely invariant with respect to the barrier position, such that no Fisher information can reach the transmission channel. As a result, all FI created in this simple scattering problem is reflected back to the source, mostly in the form of the two FI wave packets shown in the right figure of panel \textbf{b} (smaller back-reflected peaks exist, but are very small in amplitude). A video containing a numerical animation of this scattering problem is provided in Supplementary Movies M1. These observations not only demonstrate that energy and information can propagate in opposite ways; they also show that those parts of the system that depend on $\theta$ do not necessarily act as sources of information, but can also operate as information sinks.

\section{Fisher information in the near field}
To substantiate the validity of our expressions for the FI density also inside a complex scattering system (in the near field), we present an independent analysis in which a collection of $N$ atoms located near the position $\vec{r}_0$ is used as a local probe of the FI in the electromagnetic wave that can ionize these atoms. 
In analogy to the derivations provided in \cite{mandel1995optical}, where the transition rate and thus the energy transfer to the atoms is given by the energy density in the transverse part of the electric field, we provide in Supplementary Material \ref{app:FIlocaldetector} a calculation for the corresponding transfer of FI, $\CFIr_e(\para)$, to the ionisation probability of the electrons in this process. We find that, indeed, the obtained rate of FI transferred in  the ionization process is directly proportional to the FI density contained in the transverse electric field $\fidensity_{E^T}=  (\hbar \omega)^{-1} \vert \der[\theta] \Efield_\omega^T \vert^2 \epsilon_0$ in analogy to the transfer of energy, 
\begin{equation}
    \label{eq:FI_rate}
    \CFIr_e(\para) \leq 2 N \eta \fidensity_{E^T}(\vec{r}_0)\,.
\end{equation}
The equality in this relation holds if the phase of $\Efield_\omega^T$ does not change with $\para$ (e.g., by adding a reference beam with a suitable phase so that all information is contained in the intensity rather than in the phase of the field $\Efield_\omega^T$  at $\vec{r}_0$). 
The factor $\eta$ characterizes the average ionisation/detection efficiency \cite{mandel1995optical} of the atoms. Through this relation, we can assign our expression for the transversal electric part of the FI density a practical meaning also in the context of light-matter interaction. 

An interesting observation in this context is the fact that, while photodetectors can only access the transversal part of the electric field, the continuity equation \eqref{eq:cont_eq} shows that part of the FI is stored also in the longitudinal field components. These longitudinal components arise inside or in close proximity to the matter that the fields interact with, but rapidly fall off with a scaling $\propto r^{-6}$, where $r$ represents the distance to the matter (see Supplementary material \ref{supp:FI_close_proximity} for more details). Deviations from our classical analysis may arise when interactions of the above atomic detectors with the scattering system cannot be neglected anymore or a full quantum treatment becomes necessary \cite{kienesberger2023quantum}. As long as the system-detector interaction stays weak and the rotating wave approximation remains valid, we can show that the FI density of the transverse fields, integrated over all of space, is equal to the total amount of quantum FI contained in the photon field at each point in time (see Supplementary material \ref{sec:optimality} for more details).

\section{Conclusion}
In conclusion, we introduce here a novel framework to describe and evaluate the density and flow of Fisher information in electromagnetic scattering problems. In analogy to the seminal Poynting theorem that describes the conservation of energy in the propagation of waves even through highly complex environments, we formulate here a corresponding continuity equation for the creation, annihilation and conservation of Fisher information. 
Our approach 
holds significant potential for advancing the understanding of imaging systems and precision measurements. As a promising example, we mention here the field of levitated optomechanics, where great interest has recently been focused on measuring the positions of individual particles as fast and accurately as possible to cool these particles to their motional ground states \cite{tebbenjohanns2021quantum,piotrowski2023simultaneous,doi:10.1126/science.aba3993,rudolph2022force}. In this context it is thus of utmost importance to know in which direction the information on a particle's position is radiated for optimal detection 
\cite{PhysRevA.105.053504,https://doi.org/10.48550/arxiv.2212.04838}. Our approach provides the theoretical basis to quantify such FI radiation patterns for scattering particles of arbitrary shape and size, even in the presence of multiple scattering between particles. 
It is important to note that although our experimental results were specifically obtained in the microwave regime, the theoretical framework we have developed is applicable across the entire electromagnetic spectrum and can be extended to media with more complex constitutive relations. Considering also other wave equations with similar mathematical structure, our theory could serve as a framework to analyze Fisher information propagation in a broad range of different physical contexts, such as in ultrasound, seismic or quantum matter waves.

\newpage

\newpage
\section{Acknowledgments}
We thank Carlos Gonzalez-Ballestero and Oriol Romero-Isart for helpful discussions and the team behind the open-source code NGSolve for assistance. Many thanks are extended to Matthias Kühmayer for providing parts of the simulation code. Support by the Austrian Science Fund (FWF) under Project No.~P32300 (WAVELAND) is gratefully acknowledged. The computational results presented were achieved using the Vienna Scientific Cluster (VSC).

\section{Author contributions}
JH set up the theoretical framework and performed the analytical calculations under the supervision of DB, LR and SR. FR performed the experiment together with JL, under the supervision of UK. The numerical calculations were carried out together by JH, FR and LR. 
SR proposed the project. JH and SR wrote the manuscript with the assistance of FR and with input from all authors.

\clearpage
\bibliography{references} 
\clearpage

\section{Methods}
\subsection{Experiment}
\label{methods:exp}
The measurement setup and the experimental methods we present in this section are inspired by Ref.~\cite{pichler2019random}.
\subsubsection{Setup}
\label{methods:exp_setup}
\begin{figure*}
    \includegraphics[width=\textwidth]{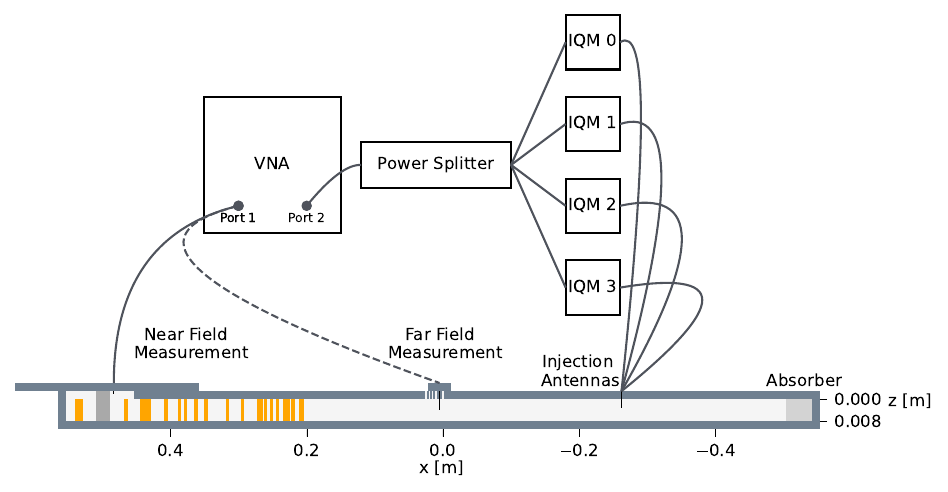}
\caption{\textbf{Experimental setup:} Cross section of the experimental setup in the $x$-$z$-plane during a far field (dotted line) and a near field (solid line) measurement. We choose the aspect ratio of the axes scaling as 1:4 ($x$-scale:$z$-scale) for better visibility. To imitate an infinitely long system, we place an absorber on the right side. On the left side, the waveguide is closed with a metallic wall. Using Teflon (orange rectangles, $8 \mm$ high) and metallic (dark gray rectangle, $11 \mm$ high) scatterers, we create a complex scattering environment on the left side. We connect the probe antenna to port 1 of the VNA and attach it to three step motors (not shown), which we use to position it. To port 2, we connect the four injection antennas via a power splitter and four IQ modulators. Note that the waveguide is higher in the near field measurement area ($11 \mm$) than in the other regions ($8 \mm$).}
\label{fig:exp_microwave_setup}
\end{figure*}

A cross section of the experimental setup is shown in Fig.~\ref{fig:exp_microwave_setup}. We work with a rectangular aluminum waveguide of length $1.103 \m$ ($x$-direction) and inner width $0.1 \m$ ($y$-direction). The system is slightly higher in the near field measurement area ($11 \mm$) than in the other regions ($8 \mm$). While the waveguide is closed with a metallic wall on the left, we use absorbing material to imitate an open lead on the right. Within the system, we randomly place 25 cylindrical Teflon scatterers of radius $2.55 \mm$ (orange rectangles). Amidst the resulting complex scattering layer, we place a cuboid metallic target scatterer of side length $2 \cm$ (dark gray rectangle). We use a pin in the bottom plate of the waveguide to fix the position of the target scatterer, which allows us to execute a precise perturbation of $\theta$.
 
To determine the microwave field for a given incoming state, we measure the transmission from port 2 to port 1 of a vector network analyzer (VNA) at different positions in the system (Note that the transmission is a unitless quantity since it is always determined relative to what we send into the system at the calibration point). Port 1 of the VNA is connected to a probe antenna, which we position using three step motors (not shown). Port 2 is connected to four injection antennas via a power splitter and four IQ modulators. Both probe and injection antennas couple weakly to the system. While the measured transmission is proportional to the $z$-component of the electric field, we can only determine the latter up to an unknown constant factor since the exact coupling strengths of the two antennas is unknown. 

The incoming state is controlled by choosing the relative phase and the attenuation of the IQ modulators (see Sec.~\ref{methods:smat_wavefrontengineering}). We can measure both the far field (via holes in the top plate) and the near field (using a movable part of the top plate with integrated antenna). As opposed to the measurement holes, the movable top plate allows us to probe the microwave field at arbitrarily close positions, enabling us to calculate the spatial derivatives in the expression for the Fisher information flux.

We measure the microwave transmission spectra in the near field measurement area on a grid of $39 \times 39$ evenly spaced points that are $2.5 \mm$ apart in $x$- and $y$-direction. Note that we could not perform measurements above the target scatterer since it touches the top plate. To evaluate the spatial gradients in the expression of the Fisher information flux on this grid, we use central finite differences. At the grid edges, we use forward/backward finite differences. The finite differences with respect to $\theta$ can be computed by performing the above measurement for the unperturbed and the perturbed system and subtracting the microwave fields. 
\subsubsection{Frequency Range and Smoothing}
\label{methods:freq_range}
 In rectangular waveguides, only $\TE{m}{n}$ and $\TM{m}{n}$ modes exist. Furthermore, each TE/TM mode has a cut-off frequency, given by
\begin{align}
\label{eq:cutoff_freq}
  \omega^2_{\text{cut-off}}(m, n) = \frac{1}{\epsilon \mu} \left( \left( \frac{n \pi}{d_y} \right)^2 + \left( \frac{m \pi}{d_z} \right)^2 \right),
\end{align}
below which they are evanescent and decay exponentially \cite{jackson1999classical}. Here $d_y=10\cm$/$d_z=8\mm$ corresponds to the extent of the waveguide in $y$-/$z$-direction. Note that for frequencies below $c/(2d_z)=18.74\GHz$, only modes with $m=0$ are non-evanescent.

We measure the transmission between port 2 and port 1 of the VNA at 2501 equally spaced frequency points in the interval $f \in [6\GHz,\ 7.5\GHz]$, ensuring that $\TE{0}{n}, n \leq 4$ are the only propagating modes. We thus avoid the $z$-dependence of the scalar microwave field $\psi \propto E_z(x,y)$. Furthermore, we can control all degrees of freedom using the four injection antennas. To reduce the measurement noise, we apply a triangular filter of width 7 to the transmission spectra, i.e., 
\begin{align*}
  T(n_f) = \frac{1}{16}&\Big(\tilde{T}(n_f-3)+ 2 \tilde{T}(n_f-2)+ 3 \tilde{T}(n_f-1)\\
  &+ 4 \tilde{T}(n_f) \\
  &+ 3 \tilde{T}(n_f+1)+2 \tilde{T}(n_f+2)+\tilde{T}(n_f+3)\Big),
\end{align*}
where $T(n_f)$/$\tilde{T}(n_f)$ denotes the smoothed/measured signal at the frequency point $n_f$, $4<n_f<2498$.

\subsubsection{Far Field Wavefunction}
\label{methods:ff_wavefunc}
\begin{figure}
\begin{tabular}{L{0.01\textwidth} F{0.48\textwidth}L{0.01\textwidth} F{0.48\textwidth}}
    \textbf{a}&
    \includegraphics[scale=0.5,valign=t]{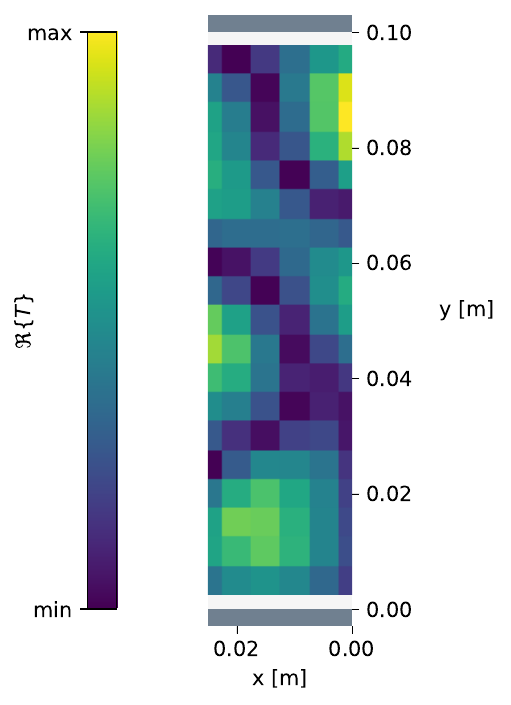}&%
    \textbf{b}&
    \includegraphics[scale=0.5,valign=t]{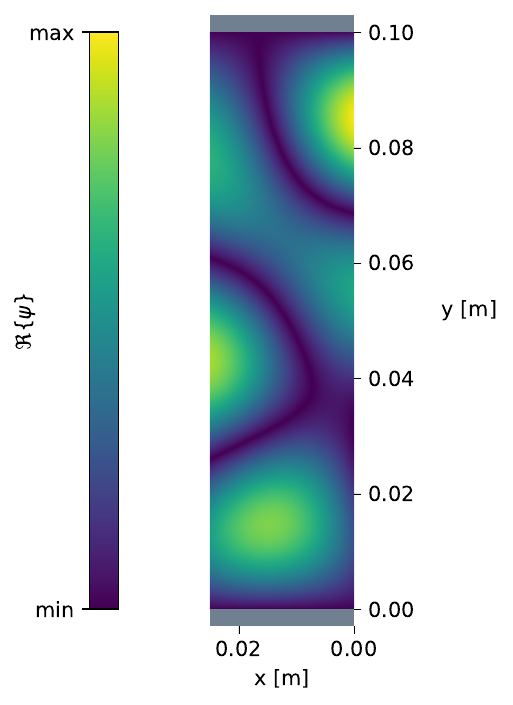}%
\end{tabular}
\caption{\textbf{Far field wavefunction:} Comparison of the measured transmission from port 2 to port 1 of the VNA at $f = 6.45 \GHz$ (\textbf{a}) and the wavefunction from Eq.~(\ref{eq:asymptwave}), where we plug in the $\cinvec{}$ and $\coutvec{}$ from the fit (\textbf{b}). We illuminate the 0th injection antenna fully while attenuating the remaining antennas maximally and show the real part of the transmission/wavefunction.} 
\label{fig:ffwavefunc}
\end{figure}

As opposed to the near field, the far field in our setup has a simple mathematical form (see Ref.~\cite{rotter2017light}):
\begin{align}
    \label{eq:asymptwave}
      \psi(x, y) = \sum_{n=1}^{N} \frac{1}{\sqrt{\kxonelead{n}}} \modeonelead{n} \left[\cinonelead{n} \expo{\imagunit \kxonelead{n} x} + \coutonelead{n} \expo{-\imagunit \kxonelead{n} x}\right],
 \end{align}
where $N = \left\lfloor \frac{d_{y} k}{\pi} \right\rfloor = 4$ denotes the number of open (flux carrying) modes, and $\cinonelead{n}$/$\coutonelead{n}$ denote the coefficients of the wavefront going in/coming out of the system. We further defined the transverse mode profiles as:
 \begin{align}
  \begin{split}
 \label{eq:transv_mode_profs}
       \modeonelead{n} = \sqrt{\frac{2}{d_{y}}} \sin{\left(\frac{n \pi y}{d_{y}}\right)}, \\
     \int_0^{d_{y}} \modeonelead{n} \modeonelead{n'} dy = \delta_{n n'}.
 \end{split}
 \end{align}
We can determine $\cinvec{}=\left(\cinonelead{1}, \cinonelead{2}, \dots, \cinonelead{N}\right)^\mathrm{T}$ and $\coutvec{}=\left(\coutonelead{1}, \coutonelead{2}, \dots, \coutonelead{N}\right)^\mathrm{T}$ at a fixed frequency and for given IQ modulator settings by measuring the microwave field in all $6 \times 19$ top-plate holes and performing a fit to Eq.~(\ref{eq:asymptwave}). More specifically, we first compute
\begin{align*}
    c_n(x) := \cinonelead{n} \expo{\imagunit \kxonelead{n} x} + \coutonelead{n} \expo{-\imagunit \kxonelead{n} x}
\end{align*}
at each $x$-position with the help of SciPy's curve\_fit function \cite{scipy}. Then, we perform a multivariate linear regression using the scikit-learn library to obtain $\cinvec{}$ and $\coutvec{}$ \cite{scikit-learn}. Fig.~\ref{fig:ffwavefunc} shows the result of such a fit for $f = 6.45 \GHz$ and the following IQM settings:
  \begin{align*}
\IQMsets{0}{0}{0}, \\
\IQMsets{i \neq 0}{40}{0},
 \end{align*}
i.e., the 0th injection antenna is fully illuminated, while the remaining antennas are on maximal attenuation. We depict the real part of the measured transmission and the far field wavefunction from Eq.~(\ref{eq:asymptwave}), where we plug in the $\cinvec{}$ and $\coutvec{}$ from the fit. Since the latter agree well, we conclude that the wave propagates freely in the far field and we can characterize it with the $\cinvec{}$ and $\coutvec{}$ from the fit.
\subsubsection{Scattering Matrix and Wavefront Engineering}
\label{methods:smat_wavefrontengineering}
The coefficients of the incoming waves $\cinvec{}$ are related to the coefficients of the outgoing waves $\coutvec{}$ via the scattering matrix $\smat$. We can thus determine the latter by measuring the $\coutvec{}$ corresponding to four linearly independent $\cinvec{}$ and by solving the system of linear equations. By illuminating the injection antennas one after the other while maximally attenuating the remaining ones, we can easily identify the required four linearly independent wavefronts. 

Since the scattering system is linear, we can inject a desired microwave state into it once we know the four linearly independent $\cinvec{}$. In this way we can produce any superposition of the latter by choosing the attenuation and the relative phase of the IQ modulators.

\subsection{Maximizing the Relative Fisher Information}
\label{app:relFI}
Here we will consider the scenario, where we gather information about the system through a receiver, while trying to avoid that this information ends up at a potential eavesdropper. More specifically, we want to find an incident field for which the ratio of the FI rates $\CFIr_R/\CFIr_E$ is maximized, where $\CFIr_R,\CFIr_E$ are the rates of FI of the receiver and eavesdropper, respectively. This problem is equivalent to maximizing the relative FI of the receiver lead $\CFIr_R/(\CFIr_E+\CFIr_R)$, which is numerically more stable and we will therefore use. To see this we can take $\CFIr_R/\CFIr_E = \lambda > 0$ and with this rewrite $\CFIr_R/(\CFIr_E+\CFIr_R)= \lambda/(1+\lambda) = f(\lambda)$, where $f(\lambda)$ is monotonically increasing for $\lambda>0$, guaranteeing that the two expressions are maximized at the same time. 
We start out by splitting the space of outgoing waves into two orthogonal subspaces of dimensions $M_R$ (at the receiver end) and $M_E$ (at the eavesdropper's end). We now use the scattering matrices $\S_R,\S_E$, which describe the scattering of the incident wave into the outgoing waves at the receiver and eavesdropper's end to define the FI operators $\matrix{F}_R,\matrix{F}_E$ on these subspaces, respectively \cite{Bouchetnatphys}. We can now reformulate the problem to
\begin{equation}
\label{eq:relFI}
    \max_{\vec{c}^{\text{in}}} \frac{[\vec{c}^{\text{in}}]^\dagger \matrix{F}_R \vec{c}^{\text{in}}}{[\vec{c}^{\text{in}}]^\dagger (\matrix{F}_R+\matrix{F}_E) \vec{c}^{\text{in}}},
\end{equation}
where the vector $\vec{c}^{\text{in}}$ indicates the incident wave. Due to the matrices being Hermitian and positive definite we can use the matrices in the denominator to define an inner product $\inner[\vec{c}][\vec{d}][] = \vec{c}^\dagger (\matrix{F}_R+\matrix{F}_E) \vec{d}$ on the incident states. We now rewrite the optimization problem ~\eqref{eq:relFI} to
\begin{equation}
\label{eq:relFI:inner}
    \max_{\vec{c}^{\text{in}}} \frac{\inner[\vec{c}^{\text{in}}][(\matrix{F}_R+\matrix{F}_E)^{-1} \matrix{F}_R \vec{c}^{\text{in}}][]}{\inner[\vec{c}^{\text{in}}][\vec{c}^{\text{in}}][]}.
\end{equation}
The operator $(\matrix{F}_R+\matrix{F}_E)^{-1} \matrix{F}_R $ is self-adjoint with respect to this inner product allowing us to apply the min-max theorem, which states that~\eqref{eq:relFI:inner} is maximized for the eigenstate corresponding to the largest eigenvalue. Due to the signal being prone to noise for waves which avoid the target, it can be beneficial to exclude waves with $[\vec{c}^{\text{in}}]^\dagger (\matrix{F}_R + \matrix{F}_E)\vec{c}^{\text{in}} <\epsilon$ for some cutoff $\epsilon$. This can be achieved by applying an orthogonal projection on both sides of the FI matrices $\matrix{F}_R,\matrix{F}_E$.

\subsection{Numerical Methods}
\label{methods:numerics}
\subsubsection{2D waveguide simulations}
\label{methods:2Dnumerics}
As described in Sec.~\ref{methods:freq_range}, only the lowest transverse mode is open between the top and the bottom plate of the experimental setup in the considered frequency range. Therefore, the $z$-component of the electric field does not depend on the $z$-coordinate, and obtain it numerically by solving the 2D Helmholtz equation 
\begin{align}
\label{eq:helmholtz}
    \left(\bigtriangleup + n^2(x, y) k_0^2\right) \psi(x, y) = 0,
\end{align}
using the finite-element library NGSolve \cite{schoberl2019ngsolve}. In Eq.~(\ref{eq:helmholtz}), $\bigtriangleup$ denotes the Laplace operator, $n(x, y)$ the position-dependent refractive index, $k_0 = \frac{2\pi}{\lambda}$ the vacuum wavenumber, and $\psi(x,y)$ the $z$-component of the electric field. We apply Dirichlet boundary conditions at the waveguide plates, and use perfectly matched layers to implement a (semi-)infinitely extended system in $x$-direction. To account for the experimental losses due to the skin effect, we add an imaginary part of $10^{-3} \ \imagunit$ to the refractive index of the entire system. 

\begin{figure}
    \centering
    \includegraphics[scale=0.75,valign=t]{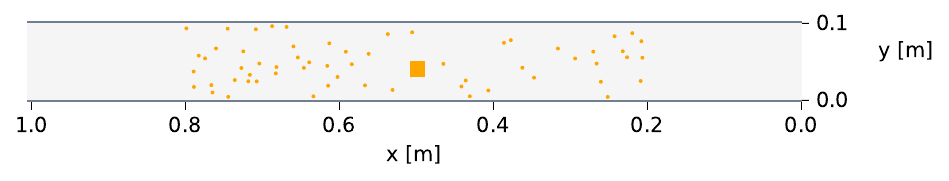}
    \caption{\textbf{Simulated 2D geometry with two open leads:} While the system is subject to Dirichlet boundary conditions at the top and the bottom side, it is infinitely long in $x$-direction. We hide a square Teflon target within a complex scattering layer of 60 circular Teflon scatterers and probe the system by injecting microwaves from the right side.}
    \label{fig:setup_2leads}
\end{figure}

We apply this framework to simulate a waveguide with two open leads in Fig.~\ref{fig:relFIflux}. While we use the 2D projection of the experimental scattering geometry on the right side of the target, we place 35 additional Teflon scatterers on its left, see Fig.~\ref{fig:setup_2leads}. We choose $2\cm$ for the side length of the square target and $2.55\mm$ for the radius of the circular scatterers in the disorder. Note that, unlike in the experiment, Teflon is used as the material of the target.

\subsubsection{1D time dependent simulations}
The time-dependent simulations are performed for scalar Maxwell equations using the finite difference method and outgoing boundary conditions. For convenience we set $c=1$ and all other units to $1$. The incident wave packet is given by 
\begin{equation}
    f(x,t) = f_0 \expo{-(t-x)^2/\tau} \sin(\omega (t-x)) 
\end{equation}
for $\tau= 0.2$ and $\omega = 500$. The scatterer barrier considered in Fig. \ref{fig:time_dependent} of the main text has a dielectric constant of $\epsilon = 2\epsilon_0$, a width of $w=1$ and the left side is located at $x=5$.

\newpage
\section{Supplementary Material}

Here we will prove the central results of this paper in detail.
In the first section we show that the maximal FI for quasi-monochromatic waves of frequency $\omega$ measured by photodetectors in the far field can be described by a FI flux. By incorporating this flux into a continuity equation we are able to identify FI sources before we consider how absorption reduces the FI. 

In section \ref{app:FIlocaldetector} and \ref{sec:optimality} we highlight that the FI density can be viewed as the local information content of the wave by first considering a weak detector in the near field and by showing that the total FI content in the photons is given by the integrated FI density in areas not in close proximity to the targets.

\subsection{Fisher information of a photodetector in the far field}
\label{app:maxFI}
We start out by showing that the FI flux for quasi-monochromatic waves is an upper bound for the Fisher information that enters a detector in the far field. The detector measures the photon flux that hits the detector surface $A$ with the mean photon flux \cite{hecht2018optics} 
\begin{equation}
    \signalflux(t) = (\hbar \omega)^{-1} \int_A \mean{\poynting}[T](t) \cdot \d \n,
\end{equation}
where $\vec{n}$ is the normal vector, which points in outgoing normal direction and  the averaged Poynting vector
\begin{equation}
    \mean{\poynting}[T](t) = T^{-1} \int_{t-T/2}^{t+T/2} \Efield(\tilde{t}) \times \Hfield(\tilde{t}) \d \tilde{t}.
\end{equation}
We assume that the noise is dominated by the shot noise and thus the signal $X$ is Poisson distributed and given by
\begin{equation}
    X(t) \sim P\left(T \signalflux(t) \right),
\end{equation}
for a photodetector measuring between $[t-T/2,t+T/2]$.
For this noise model the Fisher information for a parameter $\para$ and a single detector is given by
\begin{equation}
    \CFI(\para) = \frac{(T \der[\para] \signalflux)^2}{T \signalflux }.
\end{equation}
This can be rewritten to
\begin{equation}
\label{eq:FI_poisson}
    \CFI(\para) = \frac{(\int_{t-T/2}^{t+T/2} \int_A (\der[\para] \Efield\times \Hfield + \Efield\times \der[\para] \Hfield) \cdot \d \vec{n} \d \tilde{t})^2}{\hbar \omega \int_{t-T/2}^{t+T/2} \int_A (\Efield\times \Hfield)  \cdot \d \vec{n} \d \tilde{t}}.
\end{equation}

If we place the detector outside the scattering region we can assume that only outgoing plane waves impinge on the detector in outgoing normal direction $\n$. This gives us
\begin{equation}
\label{app:eq:FIentering}
   \int_{t-T/2}^{t+T/2} \int_A (\Efield\times \Hfield) \cdot \d \n \d \tilde{t} \geq 0
\end{equation}
for all $(\Efield,\Hfield)$.

For two such solutions $(\Efield_i,\Hfield_i)$ for $i=1,2$ we can thus define the following positive definite symmetric form 
\begin{equation}
\label{eq:inner_detector}
    \inner[(\Efield_1,\Hfield_1)][(\Efield_2,\Hfield_2)][] = \frac{1}{2 \hbar \omega }\int_{t-T/2}^{t+T/2} \int_A (\Efield_1 \times \Hfield_2 + \Efield_2 \times \Hfield_1) \cdot \d \n \d \tilde{t},
\end{equation}
where the positive definiteness follows from eq. \eqref{app:eq:FIentering}. Note that inserting the same state on both sides $\inner[(\Efield,\Hfield)][(\Efield,\Hfield)][]$ results in the average photon flow of $(\Efield,\Hfield)$ through the detector surface.
For these forms the Cauchy Schwartz inequality holds, which provides us with
\begin{equation}
\label{eq:CS_photo}
    \inner[(\Efield_1,\Hfield_1)][(\Efield_1,\Hfield_1)][] \inner[(\Efield_2,\Hfield_2)][(\Efield_2,\Hfield_2)][] \geq \left[\inner[(\Efield_1,\Hfield_1)][(\Efield_2,\Hfield_2)][]\right]^2.
\end{equation}
Equality in relation \eqref{eq:CS_photo} is satisfied for $\Efield_1 = \lambda \Efield_2$, $\Hfield_1 = \lambda \Hfield_2$ with a real scalar $\lambda$.

Finally, we note that in areas of free space the derivatives of the fields, $\der[\para] \Efield$ and $\der[\para] \Hfield$, also follow the Maxwell equations. Thus by setting $(\Efield_1,\Hfield_1)=(\Efield,\Hfield)$, $(\Efield_2,\Hfield_2)=(\der[\para]\Efield,\der[\para]\Hfield)$ and inserting eq.~\eqref{eq:CS_photo} in the FI expression (eq.~\eqref{eq:FI_poisson}) we gain an upper bound for the Fisher information
\begin{equation}
    \CFI(\para) \leq 4 (\hbar \omega)^{-1}  \int_{t-T/2}^{t+T/2} \int_A (\der[\para] \Efield\times \der[\para]\Hfield) \cdot \d \n \d \tilde{t}.
\end{equation}
Due to the additivity of independent detectors with surfaces $A_i$ this can be extended to 
\begin{equation}
    \CFI(\para) \leq 4 (\hbar \omega)^{-1}  \sum_i \int_{t-T/2}^{t+T/2} \int_{A_i} (\der[\para] \Efield\times \der[\para]\Hfield) \cdot \d \n \d \tilde{t}.
\end{equation}
For multiple detectors equality is achieved if the perturbation field components are proportional to the respective original fields at each detector surface, i.e. $\der[\para] \Efield= \lambda_i \Efield$, $\der[\para] \Hfield = \lambda_i \Hfield$ for all $A_i$ with scalars $\lambda_i$. This condition can be met if we interfere the signal with a strong reference beam matching $\der[\theta] \Efield$ at each detector (see Sec.~\ref{sec:optimal_photodetector} for a detailed consideration of ways how to implement this condition).
We will henceforth refer to the term 
\begin{equation}
\label{eq:app_fiflux}
    \fiflux(\vec{r},t) = 4 (\hbar \omega T)^{-1} \int_{t-T/2}^{t+T/2}(\der[\para] \Efield\times \der[\para]\Hfield)(\vec{r},\tilde{t}) \d \tilde{t}
\end{equation}
as the FI flux. As we are considering quasi-monochromatic waves, the frequency band of the wave is narrow around a frequency $\omega$ and can be written as
\begin{equation}
    \Efield(\r,t) = \Re [\Efield_\omega(\r,t) \expo{-\i \omega t}],
\end{equation}
where $\Efield_\omega$ corresponds to a slowly varying envelope. If we consider a detector with a measurement time $T$ that is large compared to $\omega^{-1}$ but small compared to variations in the envelope then the highly oscillating contributions get averaged out and the propagation of FI is given by its envelope. This allows us to approximate the FI flux by
\begin{equation}
\label{eq:quasiharm_flux}
	\fiflux(\vec{r},t) \approx 2 (\hbar \omega)^{-1} \Re(\der[\theta] \Efield_\omega^* \times \der[\theta] \Hfield_\omega)(\vec{r},t).
\end{equation}

\subsubsection{Optimal measurement}
\label{sec:optimal_photodetector}
In the present analysis, we derived the FI flux assuming a measurement scheme based on photodetectors located in the far field. In order for the photodetectors to reach the limit laid out by the FI flux, the information-carrying part of the wave, $\der[\theta] \Efield_\omega$, needs to match up with the rest of the outgoing field, $\Efield_\omega$. This can be achieved by interfering $\Efield_\omega$ with a strong reference field that is proportional to $\der[\para] \Efield_\omega$ at each detector surface. Here we propose ways to implement this requirement with a careful experimental design based on an array of photo detectors. 
In this context it is noteworthy to observe that the FI flux leaving the system not only sets an upper bound on the Fisher information that can be extracted by photodetectors, but on the Fisher information that can be obtained through any detection scheme of the coherent electromagnetic waves under general approximations (see Supplementary Material \ref{sec:optimality} for more details).  

We now take a reference field, where the components match the perturbed fields $\der[\theta] \Efield,\der[\theta] \Hfield$ up to a multiplicative scalar, i.e. $\der[\theta] \Efield = \lambda_i \Efield,\der[\theta] \Hfield = \lambda_i \Hfield$. The real constant $\lambda_i$ is unique for each detector surface $A_i$ over a measurement period $[0,T]$. More specifically, this implies that the reference and perturbed field are
\begin{enumerate}
    \item matched in space,
    \item matched in phase,
    \item matched in polarization,
    \item matched in time
\end{enumerate}
at each detector surface $A_i$ and interval $[0,T]$.
While these conditions seem very restrictive, we can propose ways to enforce them in practice.
In order to fulfill the first condition, we can place the detectors far enough away from the scattering region so that the outgoing waves can be approximated as plane waves. If we now match the incident angle of the perturbed and reference field with the detector surface, this results in a spatial pattern that is equal between the fields.
The phase on the other hand can either be matched by hand or by taking a random phase for the reference field, in which case the average FI degrades by a factor of $1/2$. By measuring the photons at the photodetectors for two different polarizations independently (e.g. using a polarizing beam splitter), we have a FI that is independent for each polarization allowing us to match the field for each polarization independently. Finally, the quasi-monochromatic field $\der[\theta] \Efield$ has a very slowly varying envelope (compared to the measurement time $T$), thus we only require a phase matched monochromatic reference beam at frequency $\omega$ for each measurement period $[0,T]$, which is required to be short compared to the time scale of the envelope function.

\subsection{Propagation of information}
\label{app:FIField}
In this section we will show that the Fisher information flux \eqref{eq:app_fiflux} follows a continuity equation during propagating through the system.

We start out by considering solutions $\Efield,\Hfield$ to Maxwell's equations with corresponding charges $\rho$ and currents $\J$, which depend on a parameter $\theta$. Due to linearity, the fields $\der[\para]\Efield,\der[\para]\Hfield$ solve Maxwell's equations for the charges $\der[\para]\rho$ and currents $\der[\para]\J$. The averaged Poynting vector of these fields is proportional to the FI flux, which means that its propagation is described by the averaged Poynting theorem 
\begin{equation}
\label{eq:cont_micro_app}
    \nabla \cdot \fiflux + \der[t] \fidensity_{\text{EM}} = -4 (\hbar \omega)^{-1} \mean{ \der[\para]\Efield \cdot \der[\para]\J}[T],
\end{equation}
where $\fidensity_{\text{EM}} = 2 (\hbar \omega)^{-1} \mean{\epsilon_0 (\der[\para]\Efield)^2 + \mu_0^{-1} (\der[\para]\Bfield)^2}[T] $ corresponds to the FI density, where the averages are understood as follows: $\mean{f}[T](t) = T^{-1} \int_{t-T/2}^{t+T/2} f(\tilde{t}) \d \tilde{t}$. 

\subsubsection{Linear macroscopic medium}
\label{sec:macro_media_eq}
When a wave enters a linear medium, part of the FI gets deposited in the magnetisation and polarization. However, if no losses are present then this information is extracted when it returns to free space. Thus it is convenient to include these parts in the (total) Fisher information density 
\begin{equation}
\label{eq:FI_density}
    \fidensity = 2 (\hbar \omega)^{-1} \mean{\der[\para]\Efield \cdot \matrix{\epsilon}\der[\para]\Efield +  \der[\para]\Hfield \cdot \matrix{\mu} \der[\para]\Hfield}[T],
\end{equation}
where the medium is characterized by the static tensors $\matrix{\mu},\matrix{\epsilon}$ with constitutive relations $\Dfield = \matrix{\epsilon} \Efield$ and $\Bfield = \matrix{\mu} \Hfield$.
With this we can rewrite the conservation of information in analogy to the conservation of energy as expressed by the Poynting theorem
\begin{equation}
\label{eq:cont_main_app}
    \nabla \cdot \fiflux + \der[t] \fidensity = \mean{-\der[\para]\Efield \cdot \Jeff^e - \der[\para]\Hfield \cdot \Jeff^m}[T].
\end{equation}
From this expression we can read off that the Fisher information is generated by effective electric and magnetic currents
\begin{equation}
\label{app:eq:newcurrents}
\begin{split}
        \Jeff^e = &4 (\hbar \omega)^{-1}[(\der[\para] \matrix{\epsilon}) \der[t]\Efield + \der[\para] \J_f],\\ 
        \Jeff^m =& 4 (\hbar \omega)^{-1}[(\der[\para] \matrix{\mu}) \der[t] \Hfield],
\end{split}
\end{equation}
with the current of the free charges given by $\J_f$. These effective currents are exclusively located in those areas, where matter changes with the parameter $\para$, with the interesting consequence that the Fisher information is conserved in all other areas of space. 

\subsubsection{Quasi-monochromatic waves and losses}
We will now consider systems with no free currents and an absorptive linear medium with complex scalars $\epsilon,\mu$ so that $\Dfield_\omega = \epsilon \Efield_\omega$. Associated to the FI flux given in eq.~\eqref{eq:quasiharm_flux} we now have a FI density 
\begin{equation}
\label{eq:fidensity}
    \fidensity(\vec{r},t) = (\hbar \omega)^{-1} (\vert\der[\para]\Efield_\omega \vert^2 \Re[\epsilon] +  \vert\der[\para]\Hfield_\omega\vert^2\Re[\mu])(\vec{r},t).
\end{equation}
and the FI sources
\begin{equation}
\begin{split}
	\fisource = - 2 \hbar^{-1} [\Im( \der[\theta] \Efield_\omega^{\,*} \cdot \Efield_\omega \der[\theta] \epsilon)  + \Im(\der[\theta] \Hfield_\omega^{\,*} \cdot \Hfield_\omega \der[\theta] \mu)].
\end{split}
\end{equation}
Finally, the absorption of energy contributes an additional term
\begin{equation}
	\sigma^{\text{abs}} = - 2 \hbar^{-1} [\vert\der[\theta] \Efield_\omega\vert^2 \Im(\epsilon) + \vert\der[\theta] \Hfield_\omega\vert^2 \Im(\mu) ].
\end{equation}

\subsubsection{Scattering Matrix}
We will now connect the derivation of the FI with the proof given in the main text. We consider monochromatic waves and place a photo detector in the far field of the system with the normal direction $\n$ of the detector surface pointing away from the system. This guarantees that the electromagnetic field components are orthogonal to $\n$. Then $\inner[\cdot][\cdot][] $ (given in eq.~\eqref{eq:inner_detector}) turns to an inner product allowing us to construct an orthonormal basis $\Efield_{i}$ for these states
\begin{equation}
    \Efield = \sum_i c_i^d \Efield_{i}
\end{equation}
with coefficients $c_i^d$. If we now consider an incident state with coefficients $\vec{c}^{\text{in}}$ in an arbitrary basis, the scattering matrix $\S$ connects the coefficients of the incoming and the outgoing energy flux, which arrives at the detector in the far field, $\vec{c}^d = \S \vec{c}^{\text{in}}$. 
The averaged FI flow is found by using the Hermitian form
\begin{equation}
    \frac{1}{2} \sum_i \int_{A_i} \Re(\Efield_\omega^* \times \Hfield_\omega)  \cdot \d \n_i  = T^{-1} \inner[(\Efield,\Hfield)][(\Efield,\Hfield)][] = \vec{c}^{\text{in}\,\dagger} \S^\dagger \S\vec{c}^{\text{in}},
\end{equation}
where the left side is the energy flow into the detector for the complex amplitudes $\Efield_\omega,\Hfield_\omega$.
We want to emphasize here that this relation only depends on the fields $(\Efield,\Hfield)$ in the vicinity of the detector surfaces. Thus, by letting the system medium depend on $\theta$, the waves entering the detector are still locally solutions to Maxwell's equations. Thus if the incident state does not change with $\para$, we see that
\begin{equation}
\begin{split}
    \frac{1}{2 }\sum_i \int_{A_i}\Re(\der[\theta]\Efield_\omega^* \times \der[\theta]\Hfield_\omega)  \cdot \d \n_i =& T^{-1} \inner[\der[\theta](\Efield,\Hfield)][\der[\theta](\Efield,\Hfield)][] \\
    =& \vec{c}^{\text{in}\,\dagger} \der[\theta]\S^\dagger \der[\theta]\S \vec{c}^{\text{in}}.
\end{split}
\end{equation}

\subsection{FI of a weak detector in the near field}
\label{app:FIlocaldetector}
In this section we consider a weak detector, which measures the field at any given location (even inside of the system). For this we consider the ionisation process of an atom, where the Fisher information stored in the electromagnetic field is transferred onto the ionisation process of the atom. With the source of the noise being the quantum nature of our system, we need to view the interaction between the atom and the field from a quantum mechanical perspective. If the electromagnetic field is not too weak, then the system can be described semi-classically and over long enough time scales we can further use the rotating wave approximation to describe the ionisation process of an electron. We start with a bounded electron with the corresponding ground state energy $E_0$ at time $t_0$ getting ionized toward the continuum energy $E=E_0 + \hbar \omega$ due to a quasi-monochromatic field at frequency $\omega$. The probability $p$ that a transition has taken place during the time interval $\Delta t$ is given to first order in time by (valid for $p \ll 1$) \cite{mandel1995optical}
\begin{equation}
    p \approx 2 \eta_0 \frac{u_{E^T}}{\hbar \omega} \Delta t,
\end{equation}
for the detection efficiency $\eta_0$ of the atom and the transverse electric energy density $u_{E^T} = \epsilon_0 \vert \Efield_\omega^T\vert^2/4$. It is important to note that the atom's dipole only interacts with the transverse part of the electric field. This is encapsulated in the electric dipole Hamiltonian, which only contains the transverse part of the magnetic vector potential, guaranteeing gauge invariance \cite{mandel1995optical}. 
Furthermore, the detection efficiency has a dependence on the polarisation of the light relative to the orientation of the dipole. In order to average out this degree of freedom, we now consider a system made up of $N$ atoms located in the immediate vicinity of the location $\vec{r}_0$ with random orientations. 
For each atom $i$ the ionisation probability within a time interval $[0,\Delta t]$ can be described by a Bernoulli distribution with respective probability $p_i$. We denote the ionisation probability averaged over all atoms by $\overline{p}$. 
We can now see that the rate of energy transfer $\Delta E$ from the electromagnetic field toward the electrons is proportional to the transverse electric energy density $u_{E^T}$,
\begin{equation}
    \Delta E = N \overline{p} \hbar \omega \approx 2 N \eta u_{E^T} \Delta t,
\end{equation}
where $\eta$ represents the averaged detection efficiency.

In order to derive the corresponding transfer of FI in this process, we assume that the ionisation events of the individual atoms are independent of each other. Thus we can make use of Le Cam's theorem for $\overline{p^2} \ll N^{-1}$, which tells us that we can approximate the number of measured photo-electrons within the time interval $\Delta t$ by a Poisson distribution $P(N\overline{p})$ \cite{le1960approximation}. With this we can write for the FI contained in the ionisation process in a given time interval $[0,\Delta t]$,
\begin{equation}
\label{eq:FI_ionisation}
\begin{split}
    \CFI_{e}(\para) \approx&  \frac{[N \der[\para] \overline{p}]^2}{N \overline{p}} \leq 2 N \eta  \fidensity_{E^T} \Delta t,
\end{split}
\end{equation}
where the FI density in the transverse electric field is given by $u_{E^T}^{\text{FI}} = (\hbar \omega)^{-1} \epsilon_0 \vert \der[\para]\Efield_\omega^T\vert^2$.
The inequality is due to Cauchy-Schwartz, where equality is saturated in the case where all the changes of the field with $\para$ are contained within the amplitude of the field (and not in the phase) and are thus visible in the ionisation rate (i.e., $\der[\para] \Efield_\omega^T = \lambda \Efield_\omega^T$ with a real scalar $\lambda$). The right side also corresponds to the limit that can be achieved if we measure the ionisation process of each atom independently. This shows that measuring the total count of photo-electrons is sufficient to access the overall FI from the individual ionisation processes.

Overall, eq.~\eqref{eq:FI_ionisation} shows that extraction of FI out of the field is governed by the transverse electric field part of the FI density in the same way as the transverse electric field part of the energy density determines the transfer of energy toward the electrons. This result corroborates the notion that the FI density introduced here is, indeed, a measure of the local FI content of the electromagnetic field.

\subsection{FI density in close proximity to matter}
\label{supp:FI_close_proximity}
Matter and charged particles produce longitudinal electric fields in their surroundings. Take for example a charged particle, which creates a longitudinal electric field
\begin{equation}
    \Efield^L = \frac{1}{4 \pi \epsilon_0} \frac{q}{r^2}\hat{\vec{r}},
\end{equation}
where $\hat{\vec{r}}$ is the unit vector in the direction of $\vec{r}$. While the electric field strength decays quadratically with distance, $\vert\Efield^L\vert \propto r^{-2}$, the FI density corresponding to the particle position decays much more strongly: $\fidensity_{E^L} \propto \vert \der[\para]\Efield^L\vert^2 \propto r^{-6}$. This can even be extended to more complex systems made up of matter described by a current $\J$ and a charge density $\rho$. More specifically, for the longitudinal part of the electric field we can rewrite the Ampère-Maxwell equation,
\begin{equation}
\label{eq:long_efield}
    \epsilon_0 \der[t] \Efield^L = \J^L,
\end{equation}
where the longitudinal current $\J^L$ can be written as \cite{keller2012quantum},
\begin{equation}
    \J^L(\vec{r},t) = \frac{1}{3} \J(\vec{r},t) + \frac{1}{4 \pi} P\int \frac{\matrix{1} - 3\hat{\vec{R}}^T\hat{\vec{R}}}{R^3} \J(\vec{r}',t) \d^3 \vec{r}'.
\end{equation}
Here $P$ refers to the Cauchy principal value (with spherical contraction) and $\vec{R} = \vec{r}-\vec{r}'$. 
Importantly, this means that the longitudinal current $\der[\para]\J^L$ is localized near the matter, giving us the following scaling in the distance $\tilde{r}$ to the closest part of matter:
\begin{equation}
    \der[\para]\J^L(\vec{r},t) = \mathcal{O}(\tilde{r}^{-3}).
\end{equation}
Plugging this into eq.~\eqref{eq:long_efield} provides us with the scaling for the FI density
\begin{equation}
    \fidensity_{E_\omega^L}  \propto \vert \der[\para]\Efield_\omega^L\vert^2 = \mathcal{O}(\tilde{r}^{-6}).
\end{equation}
This shows that in the semi-classical case the FI that can be extracted out of the system 
by the described photodetectors is locally given by the FI density except in areas in very close proximity to matter. On the other hand, even in these areas the FI density sets an upper bound to the extractable FI and accounts for the FI that is temporarily stored inside the longitudinal components of the electric field (e.g. in the polarisation of the matter, in the longitudinal components at the given frequency, etc.).

\subsection{Optimality of the FI density}
\label{sec:optimality}
In sections \ref{app:maxFI} and \ref{app:FIField} we have shown that the total amount of FI that can be extracted in the far field with photodetectors is given by the FI density. Here we will see that the integrated FI density not only represents a limit for photodetectors, but for any measurement device not in close proximity to matter. To this end we consider the Quantum Fisher information (QFI) $\QFI$ of a quasi-monochromatic coherent photon state, which corresponds to the maximum amount of FI that can be extracted out of a state with any measurement scheme. With the calculation of the QFI requiring a quantum description of the system, we need to quantise the electromagnetic field. The starting point is to consider a free electromagnetic field, which displays the behaviour of the QFI in regions sufficiently distant from matter. We show that the spatially integrated FI density yields the QFI of the system. Finally, we derive the same result when adding macroscopic matter and neglecting the interaction of the detector with the field and using the rotating wave approximation.

\subsubsection{QFI of the Free Electromagnetic Field}
\label{sec:free_field}
The Hamiltonian for a free transversal electromagnetic field is given by
\begin{equation}
\label{eq:quant_em}
   \hat{H}_{\text{em}} = \frac{1}{2} \int \epsilon_0 (\hat{E}^T)^2 + \mu_0^{-1} \hat{B}^2 \d \vec{r} = \int \hbar \omega(k) \left(\hat{a}_\vec{k}^\dagger \hat{a}_\vec{k} + \frac{1}{2} \right) \d \vec{k}
\end{equation}
with the transverse field operators $\hat{E}^T$ and $\hat{B}$ and the photon annihilation operators $\hat{a}_\vec{k}$ in the representation of the wave vectors $\vec{k}$, where we omitted the polarisation index for simpler notation.
As a first step, we consider a system made up of electromagnetic waves in vacuum and calculate the QFI contained in the state of the system.
In order to connect the QFI with the FI density, we introduce the operator
\begin{equation}
    \hat{K} = \int \epsilon_0 (\hat{E}^T)^- (\hat{E}^T)^+ + \mu_0^{-1} \hat{B}^- \hat{B}^+ \d \vec{r},
\end{equation}
where $(\hat{E}^T)^+/(\hat{E}^T)^-$ corresponds to the positive/negative frequency part of the transverse part of the electric field operator and correspondingly for $\hat{B}$. Using the quantisation of the electromagnetic field from eq.~\eqref{eq:quant_em}, we thus get
\begin{equation}
    \hat{K} = \int \hbar \omega(k) \hat{a}_\vec{k}^\dagger \hat{a}_\vec{k} \d \vec{k}.
\end{equation}
For a coherent quasi-monochromatic state with frequency $\omega$ given by the amplitude $\vec{\alpha}$ we thus get
\begin{equation}
   \mel{\vec{\alpha}}{\hat{K}}{\vec{\alpha}} = \frac{1}{4} \int \epsilon_0 \vert\Efield^T_\omega\vert^2 + \mu_0^{-1} \vert\Bfield_\omega\vert^2 \d \vec{r} \approx \hbar \omega \int \vert \alpha_\vec{k}\vert^2 \d \vec{k},
\end{equation}
where we used $(\hat{E}^T)^+ \ket{\vec{\alpha}} = 2^{-1} \Efield^T_\omega \expo{-\i \omega t} \ket{\vec{\alpha}}$ for coherent states and an equivalent relation for the $\hat{B}$ field operator.
If we now insert instead the coherent state given by the amplitude $\der[\para] \vec{\alpha}$ into the relation then we get
\begin{equation}
\label{eq:FI_density_QFI}
   \frac{1}{4} \int \epsilon_0 \vert\der[\para] \Efield^T_\omega\vert^2 + \mu_0^{-1} \vert\der[\para] \Bfield_\omega\vert^2 \d \vec{r} \approx \hbar \omega \int \vert \der[\para] \alpha_\vec{k}\vert^2 \d \vec{k}.
\end{equation}
To see this, we note that the quantisation of the electromagnetic field does not depend on $\para$ and thus the corresponding field solutions $\Efield_\vec{k}$ with $(\hat{E}^T)^+ = \int \Efield_\vec{k} \hat{a}_\vec{k}  \d \vec{k}$ also do not depend on $\para$. This implies that the $\para$ dependence is only contained inside the complex amplitude $\vec{\alpha}$.

Finally, we can connect the QFI for coherent states, given by $\QFI = 4 \int \vert\der[\theta] \alpha_\vec{k}\vert^2 \d \vec{k}$ and by using eq.~\eqref{eq:FI_density_QFI}, to the integrated FI density 
\begin{equation}
    \QFI(\para) = \frac{1}{\hbar \omega} \int \epsilon_0 \vert\der[\para] \Efield^T_\omega\vert^2 + \mu_0^{-1} \vert\der[\para] \Bfield_\omega\vert^2 \d \vec{r}.
\end{equation}
This shows that the integrated FI density yields the QFI at each point in time, justifying the definition of the FI density.

\subsubsection{QFI in the Presence of Matter}
The situation gets more complicated in a system with matter that interacts with the electromagnetic field. While the QFI constitutes the maximum FI that any measurement device can measure on a given state, in reality all measurement devices are also part of the physical world and thus perturb the system. While in the far field this can be neglected, in areas where the measurement device couples to the FI sources (i.e., in close proximity to the FI sources), we can expect a deviation from our theory. A recent study \cite{kienesberger2023quantum} has even suggested that this can be used to amplify the QFI for detectors in close proximity of the target. 

This discussion shows some similarities to the Abraham-Minkowski debate \cite{PhysRevLett.104.070401,Kinsler_2009} about how the momentum inside a macroscopic medium can be separated into parts associated with the field and the matter. One of the central insights in this debate has been that, depending on the experimental setup, different definitions of the field momentum have to be used. In our case, depending on the working principle of the detector, different subsystems of the full quantum state are measured (due to differences in the matter-detector coupling), resulting in differences in the QFI.

On the other hand, if we neglect this coupling and assume that the quanta of the free field Hamiltonian get measured, then we can again connect the FI density and the QFI. 
To see this connection, we take the Hamiltonian describing the full matter-field system \cite{huttner1992quantization}
\begin{equation}
    \hat{H} = \hat{H}_{\text{em}} + \hat{H}_{\text{mat}} + \hat{H}_{\text{int}}\,.
\end{equation}
Here $\hat{H}_{\text{em}}$ corresponds to the Hamiltonian of the free field, the contribution of the dressed matter part is given by the dressed matter operator $\hat{B}(\vec{k},\nu)$ (not to be confused with the magnetic field operator) such that
\begin{equation}
   \hat{H}_{\text{mat}} = \int \int_0^\infty  \hbar \nu \hat{B}^\dagger(\vec{k},\nu) \hat{B}(\vec{k},\nu) \d \nu \d \vec{k}\,,
\end{equation}
and their interaction takes on the following form
\begin{equation}
   \hat{H}_{\text{int}} = \int \int_0^\infty \hbar \xi(\vec{k},\nu) \hat{B}^\dagger(\vec{k},\nu)  (\hat{a}_{\vec{k}} + \hat{a}_{-\vec{k}}^\dagger) \d \nu \d \vec{k} + \text{H.c.}\,.
\end{equation}
Here, $\xi(\vec{k},\nu)$ describes the strength of interaction between the free field and the dressed matter \cite{huttner1992quantization}.
We now apply the rotating wave approximation on the interaction term
\begin{equation}
   \hat{H}_{\text{int}}^{\text{RWA}} = \int \int_0^\infty \hbar \xi(\vec{k},\nu) \hat{B}^\dagger(\vec{k},\nu)  \hat{a}_{\vec{k}} \d \nu \d \vec{k} + \text{H.c.}\,,
\end{equation}
to guarantee that a system starting in a coherent state stays coherent. This directly allows us to transfer the proof from the free fields (see sec. \ref{sec:free_field}) to the light-matter system. Overall, this shows that the FI density in the transversal part of the fields $u_{EM,T} = \frac{1}{\hbar \omega} \epsilon_0 \vert\der[\para] \Efield^T_\omega\vert^2 + \mu_0^{-1} \vert\der[\para] \Bfield_\omega\vert^2$ determines the QFI content of the photons. If we now add the components stored inside the polarisation and magnetisation of the matter (see sec. \ref{sec:macro_media_eq}) and in the longitudinal components of the electric field, then we arrive at the FI density given in the main text.

\end{document}